\documentclass[pra,twocolumn,aps,reprint,showpacs,superscriptaddress,amsmath,amssymb,newtxmath]{revtex4-1}
\usepackage{physics}
\usepackage[dvipdfmx]{epsfig,graphicx}
\usepackage{enumitem,bm,comment}
\usepackage{xcolor}
\usepackage[dvipdfmx,colorlinks=true,linkcolor=blue,urlcolor=blue,citecolor=blue]{hyperref}
\newcommand*{\beq}{\begin{equation}}
\newcommand*{\eeq}{\end{equation}}
\newcommand*{\beqa}{\begin{eqnarray}}
\newcommand*{\eeqa}{\end{eqnarray}}
\newcommand*{\bseq}{\begin{subequations}}
\newcommand*{\eseq}{\end{subequations}}
\newcommand*{\bal}{\begin{aligned}[b]}
\newcommand*{\eal}{\end{aligned}}
\newcommand*{\bpm}{\begin{pmatrix}}
\newcommand*{\epm}{\end{pmatrix}}
\begin{document}

\title{Superfluid properties of bright solitons in a ring}
\author{Koichiro Furutani}
\email{koichiro.furutani@phd.unipd.it}
\affiliation{Dipartimento di Fisica e Astronomia ``Galileo Galilei'', 
Universit\`a di Padova, via Marzolo 8, 35131 Padova, Italy}
\affiliation{Istituto Nazionale di Fisica Nucleare, Sezione di Padova, 
via Marzolo 8, 35131 Padova, Italy}
\author{Luca Salasnich}
\affiliation{Dipartimento di Fisica e Astronomia ``Galileo Galilei'', 
Universit\`a di Padova, via Marzolo 8, 35131 Padova, Italy}
\affiliation{Istituto Nazionale di Fisica Nucleare, Sezione di Padova, 
via Marzolo 8, 35131 Padova, Italy}
\affiliation{Istituto Nazionale di Ottica del Consiglio Nazionale delle Ricerche, 
via Carrara 2, 50019 Sesto Fiorentino, Italy}

\begin{abstract}
We theoretically investigate superfluid properties of a one-dimensional annular superfluid with a boost. 
We derive the formula of the superfluid fraction in the one-dimensional superfluid, which was originally derived by Leggett in the context of supersolid. 
We see that the superfluid fraction given by Leggett's formula detects the emergence of solitons in the one-dimensional annular superfluid. 
The formation of a bright soliton at a critical interaction strength decreases the superfluid fraction. 
At a critical boost velocity, a node appears in the soliton and the superfluid fraction vanishes. 
With a transverse dimension, the soliton alters to a more localized one and it undergoes dynamical instability at a critical transverse length. 
Consequently, the superfluid fraction decreases as one increases the length up to the critical length. 
With a potential barrier along the ring, the uniform density alters to an inhomogeneous configuration and it develops a soliton localized at one of the potential minima by increasing the interaction strength. 
\end{abstract}

\pacs{03.75.Lm; 05.45.Yv; 67.85.-d}
\maketitle

\section{Introduction}

Superfluidity is one of the most significant macroscopic quantum phenomena. 
From Kapitza's experiment with liquid helium in 1938 \cite{kapitza}, properties of superfluidity have been extensively investigated both theoretically and experimentally \cite{pethicksmith,pitaevskiistringari,svistunov}. 
For a spatial dimension lesser than three, Mermin-Wagner's theorem rules out the emergence of an off-diagonal long-range order \cite{merminwagner}. 
However, in two-dimension, we can have a quasi-long-range order and find a Berezinskii-Kosterlitz-Thouless (BKT) transition at a BKT transition temperature, above which a proliferation of free vortices occurs \cite{berezinskii,kosterlitz1,kosterlitz2,nelson}. 
This BKT transition can be observed through abrupt changes of thermodynamic quantities such as sound velocities or the superfluid fraction \cite{stringarisound,ozawa,ota,hadz,furutani}, and this is a crucial nature specific to two-dimensional systems different from three-dimensional ones. 

On the other hand, in one dimension, we can observe distinct phenomena either from two-dimensional or three-dimensional systems. 
A one-dimensional Gross-Pitaevskii (GP) equation or nonpolynomial Schr$\ddot{\text{o}}$dinger equation involves a soliton, which is a solitary wave that keeps its inhomogeneous shape during the propagation, for a sufficiently attractive interparticle interaction \cite{carr,ueda,minguzzi,polo,salasnich,crosta,amico,naldesi,naldesi20,polo22}. 
This is peculiar to effective one-dimensional systems because an extra transverse dimension can destabilize solitons in higher spatial dimensions. 
Solitons show up in various context of physics from condensed-matter physics \cite{kishine,togawa} and biological systems \cite{yomosa,scott} to high-energy physics \cite{nitta,brauner}. 
In ultracold atomic systems, the dynamics of soliton has been studied under several conditions such as the quench dynamics \cite{cidrim,muru}, the dynamics in the presence of a spin-orbit coupling \cite{tononisoc,muru}, and solitons in an annular geometry \cite{nis,polo,minguzzi}. 
Based on these developments, it is significant to clarify and characterize the property of superfluid that involves modulational changes. 

\begin{figure}[t]
\centering
\includegraphics[width=80mm]{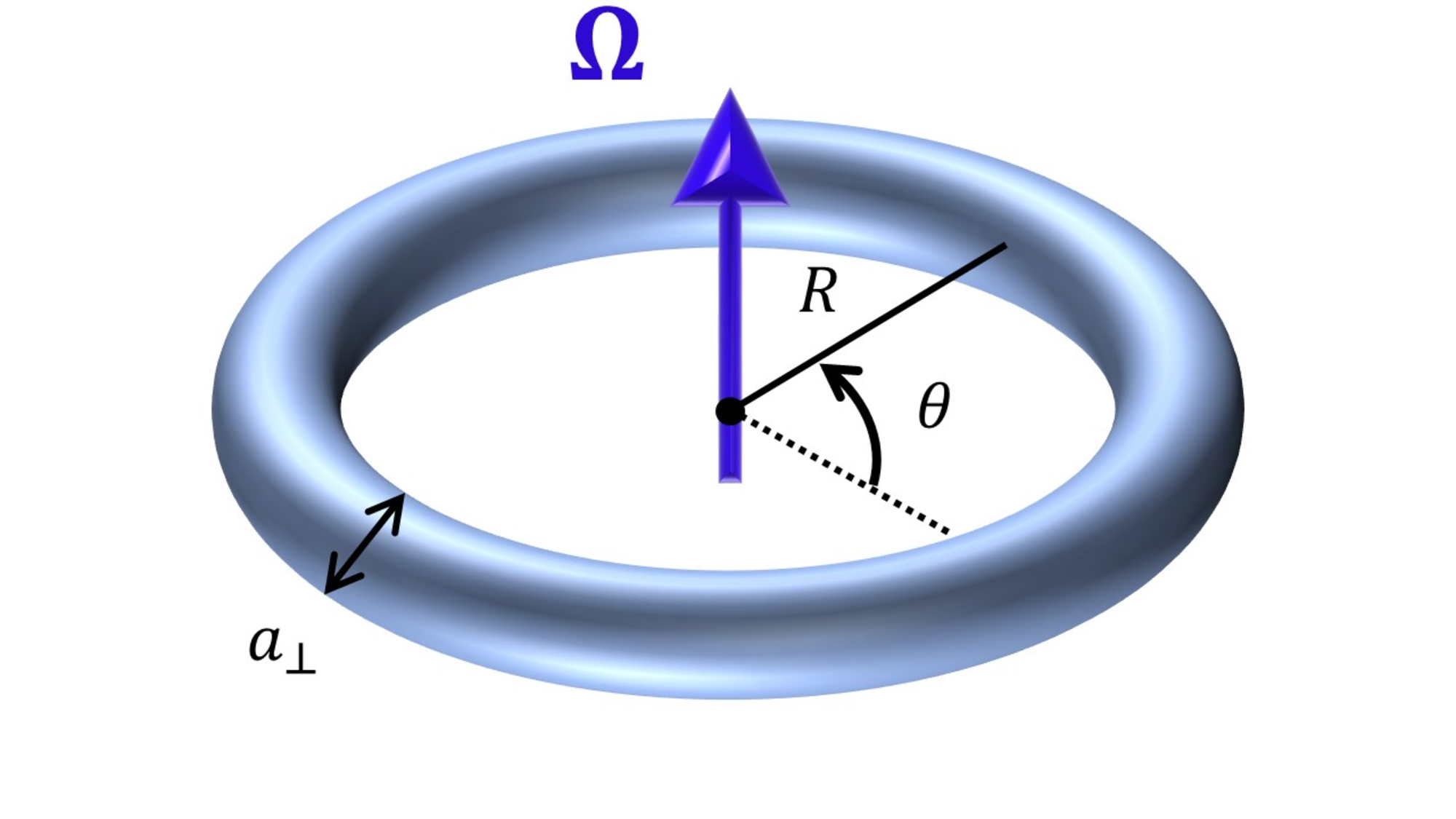}
\caption{Schematic picture of a one-dimensional superfluid in an 
annular geometry with a radius $R$ described by Eq.~\eqref{TDGP1dboost}. 
The effect of the transverse width $a_{\perp}$ is examined 
in Sec.~\ref{SecNPSE}. }
\label{BECring}
\end{figure}

In this paper, starting from the one-dimensional GP equation, we illustrate superfluid properties of the solitons in an annular geometry as depicted in Fig.~\ref{BECring} by applying Leggett's formula \cite{leggett}. 
First, we see the emergence of solitons in an annular one-dimensional superfluid with a boost. 
Secondly, we derive Leggett's formula from the one-dimensional GP equation including a boost velocity. 
While Leggett's formula is widely used mainly in the context of supersolid which involves a crystalline order associated with the nonclassical translational inertia \cite{leggett,francesca,tanzi}, we can straightforwardly obtain it also from the one-dimensional GP equation (see also \cite{martone,rica,rica94,rica07}). 
The superfluid fraction given by Leggett's formula is determined by the modulus of the macroscopic wavefunction and detects modulational instabilities in the one-dimensional superfluid. 
We see this behavior of the superfluid fraction given by Leggett's formula with varying parameters. 
As one increases the attractive interaction strength, the uniform configuration of the steady-state changes to a bright soliton at a critical interaction strength, which reduces the superfluid fraction. 
In particular, at a critical boost velocity, the soliton has a node leading to the vanishing superfluid fraction. 
Finally, we also mention the effects of the transverse dimensions and potential barriers along the ring on the superfluid fraction. 
We can see that the superfluid fraction decreases by increasing the transverse width and it undergoes dynamical instability at a certain width. 
A potential barrier makes a uniform configuration into an inhomogeneous one. 
Increasing the attractive interaction strength, we see that the inhomogeneous density develops a soliton at one of the potential minima.

\section{Bright soliton with a boost velocity}\label{Sec2}

Let us consider one-dimensional attractively interacting bosons in a ring 
with a boost velocity $u$ as in Fig.~\ref{BECring} described by
\beq
i \hbar {\partial\over\partial t} \Psi(x,t) =H_{\mathrm{GP}}[\Psi]\Psi(x,t) , 
\label{TDGP1dboost}
\eeq
with
\beq
\bal
H_{\mathrm{GP}}[\Psi]&={1\over 2m} \left(-i\hbar{\partial\over\partial x}-
mu\right)^{2} \\
&+\int^{L}_{0}dx'V(x-x')|\Psi(x',t)|^{2},
\eal
\eeq
where $\Psi(x,t)$ is the complex macroscopic wavefunction satisfying 
$\int^{L}_{0}dx\abs{\Psi(x,t)}^{2}=N$ with system size $L$, number of 
particle $N$, and $V(x)$ the interparticle interaction potential. 
Equation \eqref{TDGP1dboost} is, in the absence of the boost velocity $u=0$, 
identical to the usual GP equation. 
Now we assume a steady state
\beq 
\Psi(x,t)= \psi(x) \, e^{-i\mu t/\hbar} , 
\label{steady}
\eeq 
where $\psi(x)$ is a time-independent complex field, which describes the 
stationary configuration. 
The parameter $\mu$ is the chemical potential. This provides
\beq 
\mu \psi(x)=H_{\mathrm{GP}}[\psi]\psi(x) .
\label{s1dh}
\eeq
With the contact interaction $V(x)=g\delta(x)$, we introduce the following 
dimensionless quantities
\beq
\bal
&\theta\equiv\frac{x}{R}, \quad
\gamma\equiv\frac{2mR^{2}}{\hbar^{2}}g\bar{n}, \quad
\bar{\mu}\equiv\frac{2mR^{2}}{\hbar^{2}}\mu, \\
&\Omega\equiv \frac{mRu}{\hbar}, \quad
\varphi(\theta)\equiv \sqrt{\frac{R}{N}}\psi(\theta), 
\eal
\label{dimless}
\eeq
where $L=2\pi R$, $\bar{n}=N/L$ is the average density, $\gamma<0$ is the 
dimensionless attractive interaction strength, $\theta$ is the azimuthal 
angle, $\Omega$ is the angular frequency of rotation corresponding to the 
dimensionless boost velocity, and $\varphi(\theta)$ is the dimensionless 
macroscopic wave function normalized as 
$\int^{2\pi}_{0}d\theta\abs{\varphi(\theta)}^{2}=1$ and periodic 
$\varphi(0)=\varphi(2\pi)$. 
Then, Eq.~\eqref{s1dh} can be written as
\beq
\bar{\mu}\varphi(\theta)=\left[\left(-i\frac{\partial}{\partial\theta}
-\Omega\right)^{2}+2\pi\gamma\abs{\varphi(\theta)}^{2}\right]\varphi(\theta).
\eeq
The GP equation has a bright soliton solution as a stationary 
solution given by \cite{carr,ueda}
\beq
\varphi(\theta)=
\begin{cases}
\displaystyle\sqrt{\frac{1}{2\pi}} & \text{($\abs{\gamma}\le
\abs{\gamma_{\mathrm{c}}}$)}, \\
\displaystyle\sqrt{\frac{K(m)}{2\pi E(m)}}\mathrm{dn}\left(\frac{K(m)}{\pi}
\left(\theta-\theta_{0}\right)\Big|m\right) & 
\text{($\abs{\gamma}>\abs{\gamma_{\mathrm{c}}}$)},
\end{cases}
\label{analyticGPrest}
\eeq
where $K(m)$ and $E(m)$ are the first and second kind complete elliptic 
integrals, respectively, under the condition $K(m)E(m)=\pi^{2}\gamma/2$ 
and $\mathrm{dn}(u|m)$ is a Jacobi elliptic function with an elliptic modulus $m$. 
Here $\gamma_{\mathrm{c}}=-1/2$ is the critical interaction strength. 
Due to the broken continuous translational symmetry, 
Eq.~\eqref{analyticGPrest} includes a parameter $\theta_{0}$ that specifies 
the center of the bright soliton. 
This translational symmetry breaking results in the emergence of a 
Nambu-Goldstone (NG) mode in addition to the one associated 
with the $\mathrm{U}(1)$ symmetry breaking \cite{ueda,kanamoto,martone}. 
This emergence of two distinct NG modes are analogous to {\it supersolid}, 
which has both a crystalline order and a superfluid order 
\cite{leggett,rica,rica07,martone,stringari,watanabe,rica94,tanzi}. 
In the following, we set $\theta_{0}=0$ for brevity. 
It is important to stress that the NG mode related 
to the formation of the bright soliton breaking the continuous 
translational symmetry is gapless, i.e. zero energy with respect to the ground state. 
The first excitation energy calculated under the Bogoliubov approximation is gapped at $\gamma=0$, while it closes at $\gamma=\gamma_{\mathrm{c}}$ \cite{ueda}. 
The latter phenomenon is also usually called modulational instability. 
Indeed, the Bogoliubov spectrum of the uniform configuration 
reads $E_l^{(\mathrm{B})}=\sqrt{l^2(l^2+2\gamma)}$ 
with $l=\pm 1,\pm 2,\pm 3,...$ \cite{ueda,kanamoto} and for $l=\pm 1$ one gets 
$E^{(\mathrm{B})}=0$ precisely at $\gamma=\gamma_{\mathrm{c}}=-1/2$. 
Note that at $\gamma=0$ our bosonic 
system is noninteracting and uniform but also 
fully superfluid just because there is a finite energy gap 
due to the finite size. 

With a boost velocity, we can define phase winding 
number $\nu$, which is an integer such that $\Omega-1/2<\nu\le\Omega+1/2$, 
and the relative angular frequency 
\beq
\omega\equiv\Omega-\nu ,
\eeq
which varies within $-1/2\le\omega<1/2$ \cite{kanamoto,ueda08,ueda10}. 
Below the critical interaction strength $\gamma<\gamma_{\mathrm{c}}
\equiv2\omega^{2}-1/2$, it has a bright soliton. 
The number density $n(\theta)=\abs{\varphi(\theta)}^{2}$ is \cite{kanamoto}
\beq
n(\theta)=
\begin{cases}
\displaystyle\mathcal{N}^{2}\left[\mathrm{dn}^{2}\left(\frac{K(m)}{\pi}
\theta\Big|m\right)-\eta m'\right] 
& \text{\Big($0<\abs{\omega}<\dfrac12$\Big)}, \\
\displaystyle\mathcal{N}^{2}\mathrm{dn}^{2}\left(\frac{K(m)}{\pi}
\theta\Big|m\right) 
& \text{($\omega=0$)}, \\
\displaystyle\mathcal{N}^{2}m\abs{\mathrm{cn}\left(\frac{K(m)}{\pi}
\theta\Big|m\right)}^{2} 
& \text{\Big($\omega=-\dfrac12$\Big)}, \\
\end{cases}
\label{ntheta}
\eeq
where $m'=1-m$ and
\beq
\mathcal{N}^{2}=\frac{K(m)^{2}}{\pi^{3}\abs{\gamma}} ,\quad 
\eta=\frac{f_{\mathrm{d}}}{2m'K(m)^{2}}=1-\frac{f_{\mathrm{c}}}{2m'K(m)^{2}},
\eeq
\bseq
\beq
f\equiv2K(m)^{2}-2K(m)E(m)-\pi^{2}\gamma,
\eeq
\beq
f_{\mathrm{c}}\equiv2m'K(m)^{2}-2K(m)E(m)-\pi^{2}\gamma,
\eeq
\beq
f_{\mathrm{d}}\equiv2K(m)E(m)+\pi^{2}\gamma.
\eeq
\eseq
Equation \eqref{ntheta} indicates that, for $0\le\abs{\omega}<1/2$, 
the number density is nodeless while it has a node for $\omega=-1/2$. 
The chemical potential $\bar{\mu}$ is determined by \cite{kanamoto}
\beq
\bar{\mu}=
\begin{cases}
\displaystyle\frac{f_{\mathrm{d}}-f_{\mathrm{c}}-f}{2\pi^{2}}
& \text{\Big($0<\abs{\omega}<\dfrac12$\Big)}, \\
\displaystyle-\frac{K(m)^{2}}{\pi^{2}}(1+m')
& \text{($\omega=0$)}, \\
\displaystyle-\frac{K(m)^{2}}{\pi^{2}}(1-2m')
& \text{\Big($\omega=-\dfrac12$\Big)}, \\
\end{cases}
\eeq
and the elliptic modulus $m$ is determined by
\bseq
\beq
2\pi\abs{\omega}=\sqrt{\frac{2f_{\mathrm{d}}f_{\mathrm{c}}}{f}}+
\pi\left[1-\Lambda_{0}(\varepsilon|m)\right] \quad 
\text{\Big($0<\abs{\omega}<\dfrac12$\Big)}, 
\eeq
\beq
f_{\mathrm{d}}=0 \quad \text{($\omega=0$)},
\eeq
\beq
f_{\mathrm{c}}=0 \quad \text{\Big($\omega=-\dfrac12$\Big)},
\eeq
\eseq
with $\varepsilon\equiv\mathrm{arcsin}\left(\sqrt{f_{\mathrm{c}}/(m'f)}\right)$ 
and
\beq
\Lambda_{0}(\varepsilon|m)
\equiv\frac{2}{\pi}\left[K(m)\mathcal{E}(\varepsilon|m')-\left[K(m)-E(m)
\right]F(\varepsilon|m')\right].
\eeq
Here, $F(u|m)$ and $\mathcal{E}(u|m)$ are the elliptic integrals of the first kind and second kind respectively. 

\section{Superfluid fraction associated with nonclassical translational inertia}

Let us now set 
\beq 
\psi(x) = n(x)^{1/2} \, e^{i \phi(x)} 
\label{decompose}
\eeq
and 
\beq 
v(x) = {\hbar\over m}{\partial\over \partial x} \phi(x) \; . 
\label{vcrucial}
\eeq
Inserting these formulas into Eq.~\eqref{s1dh} we get 
\beq
\bal
\left[-{\hbar^2\over 2m} {\partial^2\over\partial x^2} 
+ {m\over 2}\left[v(x)-u\right]^{2} + gn(x) \right] 
n(x)^{1/2} = \mu n(x)^{1/2}, 
\eal
\label{q1}
\eeq
\beq
{\partial \over \partial x}\left[ n(x) 
\left( v(x) - u \right) \right] = 0 \; . 
\label{q2}
\eeq
Equation \eqref{q2} implies that 
\beq 
n(x) \left( v(x) - u \right) = J \; , 
\label{current}
\eeq
where $J$ is a constant current density. 
Equation \eqref{current} is very interesting because it says that 
if $n(x)$ has spatial variations then also $v(x)$ must have spatial 
variations. 

We now introduce the average value of the velocity $v(x)$ in a 
spatial region $[a,b]$ as 
\beq 
{\bar v} = {1\over b-a} \int_{a}^b v(x) \, dx \; ,  
\label{media}
\eeq 
Then, from Eqs.~\eqref{current} and \eqref{media} we obtain 
\beq 
{\bar v} = {1\over b-a} 
\int_{a}^b \left( {J\over n(x)} + u \right) 
\, dx = {J\over {\bar n}_{\mathrm{s}}} + u  \; , 
\eeq
where 
\beq 
{\bar n}_{\mathrm{s}} = \left[{1\over b-a} \int_a^b {dx\over n(x)}\right]^{-1} \; . 
\label{super-leggett}
\eeq
We will see that the number density ${\bar n}_{\mathrm{s}}$ can be interpreted 
as the superfluid number density of the stationary state 
in the spatial region $[a,b]$. 
Indeed, Eq.~\eqref{super-leggett} 
is the 1D version of the formula obtained by Leggett 
in 1970 \cite{leggett} for a supersolid with spatial periodicity $b-a$, 
and recently discussed by many others \cite{rica,chomaz,martone}. 
In other words, if the stationary state $\psi(x)$ 
moves with the average velocity ${\bar v}$, its current density reads 
\beq 
J = {\bar n}_{\mathrm{s}} 
\left( {\bar v} - u\right) \; , 
\label{current1}
\eeq
where ${\bar v}$ is the average velocity in the region $[a,b]$
and ${\bar n}_{s}$ the corresponding superfluid number density. 

Let us now consider the linear canonical momentum in the region $[a,b]$. 
It is given by 
\beq 
{\cal P} = -{i\hbar\over 2} \int_a^b \left[ 
\psi^*(x) {\partial\over\partial x} \psi(x) 
- \psi(x) {\partial\over\partial x} \psi^*(x) \right] \, dx \; . 
\label{lineP}
\eeq
Note that Eq.~\eqref{lineP} is different from the physical momentum, which takes into account the boost velocity, $P=\mathrm{Re}\int^{b}_{a}dx\psi^{*}(x)\left[-i\hbar\partial_{x}-mu\right]\psi(x)$. 
With the help of Eqs.~\eqref{decompose} and \eqref{current}, we then find 
\beq
\bal
{\cal P} &= 
m \int_a^b n(x) \, v(x) \, dx = 
m \int_a^b \left( J + n(x) u \right) dx \\
&=m \left(J + {\bar n} u \right) (b-a) \\
&= m \left[  {\bar n}_{\mathrm{s}} 
\left( {\bar v} - u\right) +  {\bar n} u \right] (b-a) \; ,  
\eal
\label{piopaio}
\eeq
where 
\beq 
{\bar n} = {1\over b-a} \int_a^b n(x) \, dx 
\eeq
is the average number density in the region $[a,b]$. 

As usual (see for instance Refs.~\cite{leggett,rica}), 
the normal density ${\bar n}_{\mathrm{n}}$ in 
the spatial region $[a,b]$ can be defined as the response of the 
linear momentum ${\cal P}$ to the boost velocity $u$, namely 
\beq 
{\bar n}_{\mathrm{n}} = 
{1\over m (b-a)}{\partial {\cal P} \over \partial u} \; .  
\eeq
By using Eq.~\eqref{piopaio}, we immediately find 
\beq
{1\over m (b-a)}{\partial {\cal P} \over \partial u} 
= {\bar n} - {\bar n}_{\mathrm{s}} \; ,  
\eeq
and this result fully justifies that ${\bar n}_{\mathrm{s}}$ of Eq.~\eqref{super-leggett} 
is indeed the superfluid number density in the region $[a,b]$. 
Similarly, a relation with the nonclassical translational inertia (NCTI), 
which is expected in the case of a supersolid system, is provided by the
definition of superfluid fraction $f_{\mathrm{s}}$ associated with NCTI \cite{rica}
\beq
f_{\mathrm{s}}
=\lim_{u \to 0} \left[ 
1-\frac{1}{Nm} {\partial {\cal P}\over \partial u}\right] ,
\eeq
with $N=\int_a^b n(x) \, dx=(b-a){\bar n}$. 
Using the above definition for $\cal{P}$, one finds
\beq
\bal
f_{\mathrm{s}}=\frac{{\bar n}_{\mathrm{s}}}{{\bar n}} &= \left[{N\over (b-a)^2}
\int_a^b {dx\over n(x)} \right]^{-1} \\
&=(b-a)^{2}\left[\int^{b/R}_{a/R}\frac{d\theta}{\abs{\varphi(\theta)}^{2}}
\right]^{-1}. 
\eal
\eeq
In particular, by setting $a=0$ and $b=L=2\pi R$, we can obtain
\beq
\bal
f_{\mathrm{s}}&=\left[\frac{1}{L}\int^{L}_{0}dxn(x)\cdot\frac{1}{L}
\int^{L}_{0}\frac{dy}{n(y)}\right]^{-1} \\
&=4\pi^{2}\left[\int^{2\pi}_{0}\frac{d\theta}{\abs{\varphi(\theta)}^{2}}
\right]^{-1},
\eal
\label{leggettformula}
\eeq
where $n(x)=\abs{\psi(x)}^{2}$ is the number density.

\subsection{Superfluid fraction in the strong-coupling regime at rest}

\begin{figure}[t]
\centering
\includegraphics[width=80mm]{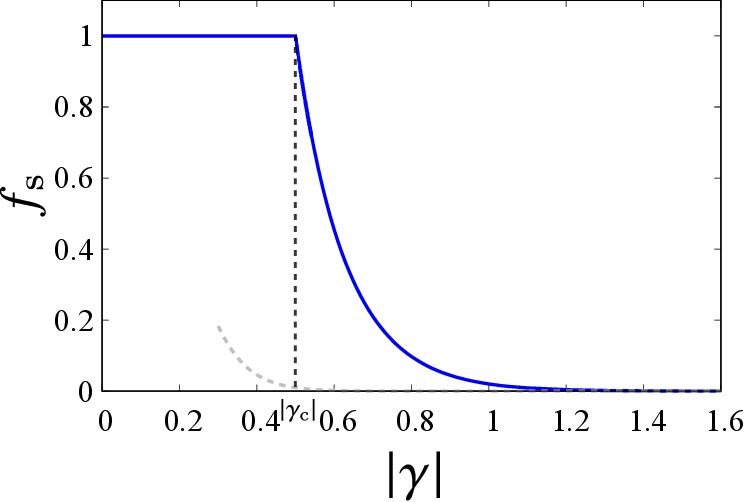}
\caption{Superfluid fraction calculated by Eqs.~\eqref{analyticGPrest} 
and \eqref{leggettformula}. The gray dashed curve represents the approximated 
result of Eq.~\eqref{leggettstrong} in the strong-coupling regime 
$\abs{\gamma}\gg\abs{\gamma_{\mathrm{c}}}$.}
\label{sfracbs}
\end{figure}

In the strong-coupling regime $\gamma\gg\gamma_{\mathrm{c}}$ without boost, 
one can write \cite{ueda}
\beq
\varphi(\theta)\simeq \sqrt{\frac{\pi\abs{\gamma}}{4}}
\sech{\left(\frac{\pi\abs{\gamma}}{2}\theta\right)} ,
\eeq
and the superfluid fraction given by Leggett's formula can be 
calculated analytically as \cite{zakharov}
\beq
f_{\mathrm{s}}
=\alpha^{2}\coth{\left(\alpha\right)}\left[\frac{\alpha}{2}+\frac14 
\sinh{\left(2\alpha\right)}\right]^{-1} ,
\label{leggettstrong}
\eeq
with $\alpha\equiv \pi^{2}\abs{\gamma}$.

The superfluid fraction calculated by Eq.~\eqref{leggettformula} is 
displayed in Fig.~\ref{sfracbs}. 
The blue solid line shows the result by Eq.~\eqref{analyticGPrest} while the 
gray dashed line stands for the analytic result of Eq.~\eqref{leggettstrong} in the strong-coupling regime. 
We numerically calculated the stationary modulus from Eq.~\eqref{TDGP1dboost} using the Crank-Nicolson method. 
In the uniform regime $\abs{\gamma}\le\abs{\gamma_{\mathrm{c}}}$, the superfluid fraction remains $f_{\mathrm{s}}=1$. 
At the critical interaction strength $\gamma=\gamma_{\mathrm{c}}$, it sharply drops, and in the solitonic regime $\abs{\gamma}>\abs{\gamma_{\mathrm{c}}}$, it monotonically decreases. 
The inhomogeneous density in the solitonic regime leads to the non-quantized angular momentum unlike the uniform case \cite{kanamoto}. 
It means that the angular momentum is no longer a good quantum number in the solitonic regime. 
However, the circulation of superfluid velocity is still quantized as \cite{kanamoto}
\beq
\int^{2\pi}_{0}\frac{d\theta}{2\pi}\partial_{\theta}\phi(\theta)=\frac{1}{2\pi}\left[\phi(2\pi)-\phi(0)\right]
=\nu,
\label{wind}
\eeq
and therefore the superfluidity is sustained. 

One may wonder if $f_{\mathrm{s}}=1$ in the noninteracting case $\gamma=0$ contradicts Landau's criterion. 
Note that Leggett's formula \eqref{leggettformula} focuses only on the ground state instead of including the effects of elementary excitations. 
In a noninteracting Bose gas, Landau's criterion indeed rules out superfluidity because of the elementary excitations and the superfluid fraction given by Landau's formula within the two-fluid model vanishes. 
In an infinite size system, the Bogoliubov spectrum is continuum and gapless. 
It can make the superfluid ground state fragile to elementary excitations based on Landau's criterion. 
In our finite-size system, however, the Bogoliubov spectrum is discretized and gapped except for $\gamma=\gamma_{\mathrm{c}}$ \cite{ueda,kanamoto}. 
Consequently, the superfluid ground state can be stabilized, and we have $f_{\mathrm{s}}=1$ even in the noninteracting case. 

\begin{figure}[t]
\centering
\includegraphics[width=80mm]{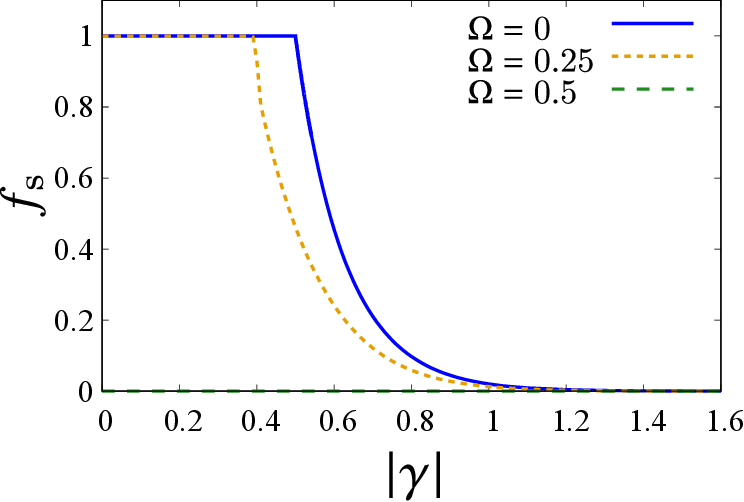}
\caption{Interaction dependence on superfluid fraction with boost 
velocities $\Omega=0, 0.25, 0.5$. 
The curves are numerically computed from Eqs.~\eqref{TDGP1dboost} and \eqref{leggettformula}.}
\label{sfracgam}
\end{figure}

\begin{figure}[t]
\centering
\includegraphics[width=80mm]{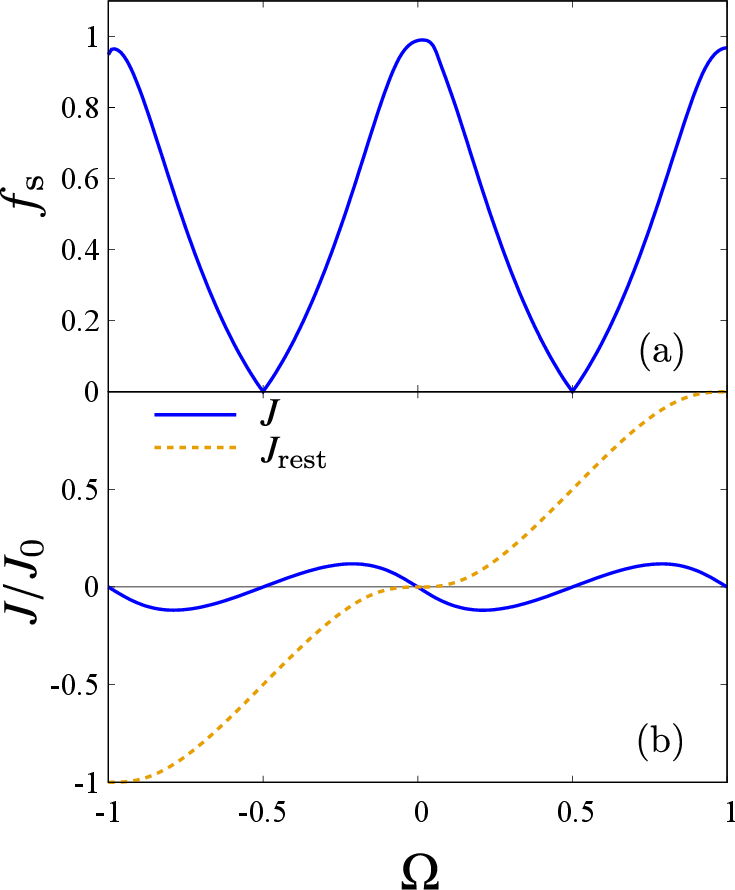}
\caption{Superfluid fraction and current density scaled by $J_{0}=\hbar\bar{n}/(mR)$ as functions of the dimensionless boost velocity $\Omega$ with a fixed interaction strength $\gamma=-1/2$. 
They are numerically calculated from Eqs.~\eqref{TDGP1dboost} and \eqref{leggettformula}. }
\label{sfracomega}
\end{figure}

\subsection{Superfluid fraction with a boost velocity}

In the presence of the boost velocity $u$, using the formula of Jacobi's elliptic functions
\beq
\int\frac{dx}{\mathrm{dn}^{2}(x|m)}
=\frac{1}{m'}\left[\mathcal{E}(x|m)-m\frac{\mathrm{cn}(x|m)\mathrm{sn}(x|m)}{\mathrm{dn}(x|m)}\right] ,
\eeq
\beq
\bal
&\int\frac{dx}{\mathrm{dn}^{2}(x|m)-A} \\
&= \Pi\left(\frac{m}{1-A};\mathrm{am}(x|m)\Big|m\right)\frac{\mathrm{dn}(x|m)}{(1-A)\sqrt{1-m\mathrm{sn}^{2}(x|m)}},
\eal
\eeq
with $A$ a real constant and $\Pi(n;x|m)$ the elliptic integral of the third kind, one obtains the superfluid fraction
\beq
f_{\mathrm{s}}=
\begin{cases}
\displaystyle\frac{\mathcal{N}^{2}}{\bar{n}}\frac{f}{2K(m)}\Pi\left(\frac{2mK^{2}(m)}{f}\Big|m\right)
& \text{\Big($0<\abs{\omega}<\dfrac12$\Big)}, \\
\displaystyle\frac{\mathcal{N}^{2}}{\bar{n}}\frac{m'K(m)}{E(m)} ,
& \text{($\omega=0$)}, \\
\displaystyle 0
& \text{\Big($\omega=-\dfrac12$\Big)}, \\
\end{cases}
\label{fsrot}
\eeq
where $\Pi(n|m)\equiv\Pi(n;\pi/2|m)$ is the complete elliptic integral of the third kind.

Figure \ref{sfracgam} shows the superfluid fraction numerically computed from Leggett's formula and Eq.~\eqref{TDGP1dboost} with some values of the boost velocity. 
For a superfluid at rest $\Omega=0$, the critical interaction strength is $\gamma_{\mathrm{c}}=-1/2$ and the superfluid fraction monotonically decreases as one increases the attractive interaction strength. 
With a boost velocity, the critical interaction strength is $\gamma_{\mathrm{c}}=2\omega^{2}-1/2$ and the superfluid fraction decreases for $\abs{\gamma}>\abs{\gamma_{\mathrm{c}}}$ as well. 
In particular, for $\Omega=-1/2$, the relative angular frequency is $\omega=-1/2$ and the number density has a node. 
As a result, the superfluid fraction computed by Leggett's formula vanishes.

\begin{figure*}[t]
\begin{tabular}{cc}
\begin{minipage}{0.5\hsize}
\centering
\includegraphics[scale=0.7]{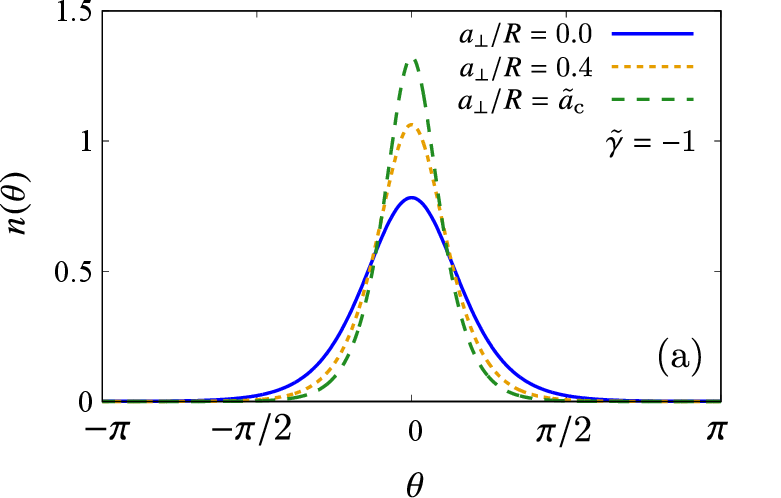}
\end{minipage}
\begin{minipage}{0.5\hsize}
\centering
\includegraphics[scale=0.7]{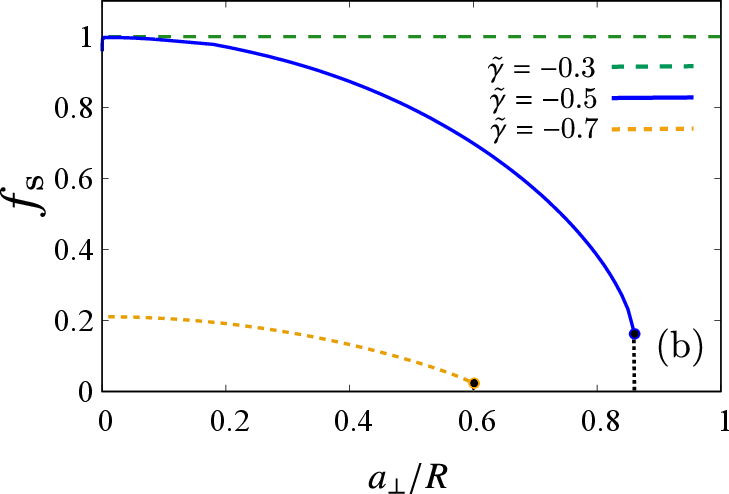}
\end{minipage}
\end{tabular}
\caption{Density distribution $n(\theta)=\abs{\varphi(\theta)}^{2}$ for $\Tilde{\gamma}=-1$ and the superfluid fraction in the presence of a transverse width $a_{\perp}$ and the superfluid fraction as a function of the characteristic transverse length $a_{\perp}$ without a boost velocity $\Omega=0$. 
The blue solid curve in the left panel (a) represents the configuration in the absence of the transverse width. 
The orange dotted curve represents the result for $\Tilde{a}_{\perp}=0.4$. 
The green dashed curve stands for the result for $\Tilde{a}_{\perp}=\Tilde{a}_{\mathrm{c}}$, above which the system is dynamically unstable. 
The superfluid fraction in the right panel (b) is numerically calculated from Eqs.~\eqref{1dNPSEdimless} and \eqref{leggettformula}. 
The endpoints correspond to the critical transverse width $\Tilde{a}_{\mathrm{c}}$ beyond which the bright soliton collapses.}
\label{densfracap}
\end{figure*}

In Fig.~\ref{sfracomega}, we show the angular frequency dependence of superfluid fraction in the upper panel (a) and current density in the lower panel (b) with $\gamma=-1/2$ numerically calculated from Eq.~\eqref{TDGP1dboost}. 
As in Ref.~\cite{kanamoto}, the number density is periodic in $\Omega$. Consequently, the superfluid fraction also exhibits periodicity.
The superfluid fraction decreases as one rotates the superfluid and, as mentioned, it vanishes at $\Omega=\pm 1/2$ corresponding to $\omega=-1/2$. 
By virtue of Eq.~\eqref{wind}, the current density in Eq.~\eqref{current1} can be written as
\beq
\frac{J}{J_{0}}=-f_{\mathrm{s}}\omega ,
\label{Jrot}
\eeq
in the corotating frame with $J_{0}\equiv\hbar\bar{n}/(mR)$, or, in the rest frame, 
\beq
\bal
\frac{J_{\text{rest}}}{J_{0}}
&=\frac{J}{J_{0}}+\Omega \\
&=f_{\mathrm{n}}\Omega+f_{\mathrm{s}}\nu ,
\eal
\label{Jrest}
\eeq
with $f_{\mathrm{n}}\equiv1-f_{\mathrm{s}}$ the normal fluid fraction. 
The current densities given by Eqs.~\eqref{Jrot} and \eqref{Jrest} are displayed in Fig.~\ref{sfracomega}(b). 
The current density in the corotating frame $J$ oscillates in $\Omega$. 
For $\Omega$ an integer, the relative angular frequency $\omega$ vanishes and $J$ vanishes as well. 
For $\Omega$ a half-integer, the superfluid fraction vanishes and $J$ vanishes as well. 
Consequently, $J$ has nodes under the angular frequencies at which $\Omega$ is an integer or a half-integer. 
It is distinct from the superfluid fraction in Fig.~\ref{sfracomega}(a) that the current density in the corotating frame is odd in $\Omega$ because it involves the relative angular frequency as well as the superfluid fraction. 

Besides the exact solution in Eq.~\eqref{ntheta}, we can easily understand the presence of a node in the wave function for $\omega=-1/2$ through a unitary transformation. 
We can write the GP Hamiltonian with a boost velocity as
\beq
\bal
H_{\mathrm{GP}}(\Omega)&=\frac{1}{2m}\left(-i\hbar\frac{\partial}{\partial x}-\frac{\hbar\Omega}{R}\right)^{2}+g\abs{\Psi(x,t)}^{2} \\
&=e^{i\Omega x/R}H_{\mathrm{GP}}(0)e^{-i\Omega x/R} .
\eal
\eeq
Using $H_{\mathrm{GP}}(0)$, we can write the GP equation as
\beq
i\hbar\frac{\partial}{\partial t}\Tilde{\Psi}(x,t)=\left[H_{\mathrm{GP}}(0)+\hbar\dot{\Omega}\frac{x}{R}\right]\Tilde{\Psi}(x,t) ,
\eeq
with
\beq
\Tilde{\Psi}(x,t)\equiv e^{-i\Omega x/R}\Psi(x,t) .
\eeq
In particular, for a constant boost velocity $\Omega(t)=\Omega$ and assuming the steady solution in Eq.~\eqref{steady}, one obtains
\beq
\Tilde{\varphi}(\theta)=e^{-i\Omega\theta}\varphi(\theta) ,
\eeq
which leads to a twisted boundary for $\Tilde{\varphi}(\theta)$. 
Under the single value condition $\varphi(2\pi)=\varphi(0)$, for $\Omega\in\mathbb{Z}$, the transformed one also satisfies $\Tilde{\varphi}(2\pi)=\Tilde{\varphi}(0)$. 
Hence, the number density and superfluid fraction remain the same as the case without a boost velocity $\Omega=0$. 
In the case of half integer $\Omega=n+1/2$, on the other hand, one obtains $\Tilde{\varphi}(2\pi)=-\Tilde{\varphi}(0)$. 
This change of sign requires at least one node in $\Tilde{\varphi}(\theta)$.

\section{Effects of transverse dimensions}\label{SecNPSE}

A realistic 1D system would have a finite width in the transverse direction as in Fig.~\ref{BECring}. 
This effect can be captured by the 1D nonpolynomial Schr$\mathrm{\ddot{o}}$dinger (NPS) equation \cite{salasnicha,salasnichb,adhikari,salasnich}. 
It is given by
\beq
i\hbar\frac{\partial}{\partial t}\Psi(x,t)=H_{\mathrm{NPS}}[\Psi]\Psi(x,t), 
\label{TDNPS}
\eeq
with
\beq
\bal
H_{\mathrm{NPS}}[\Psi]&=\frac{1}{2m}\left(-i\hbar\frac{\partial}{\partial x}-mu\right)^{2}+\Tilde{g}\frac{\abs{\Psi(x,t)}^{2}}{\eta^{2}} \\
&+\frac{\hbar\omega_{\perp}}{2}\left(\frac{1}{\eta^{2}}+\eta^{2}\right) ,
\eal
\eeq
with $\Tilde{g}\equiv g_{\mathrm{3D}}/(2\pi a_{\perp}^{2})$, $g_{\mathrm{3D}}=4\pi\hbar^{2}a_{s}/m$,  $a_{\perp}=\sqrt{\hbar/(m\omega_{\perp})}$ the transverse width, and $\eta\equiv[1+2a_{s}\abs{\Psi(x,t)}^{2}]^{1/4}$ \cite{salasnicha,salasnichb}. 
The steady-state equation can be found under Eq.~\eqref{steady} as
\beq
\mu\psi(x)=H_{\mathrm{NPS}}[\psi]\psi(x). 
\eeq
Using the dimensionless quantities in Eq.~\eqref{dimless}, one obtains
\beq
\bal
\bar{\mu}\varphi(\theta)
&=\Bigg[\left(-i\frac{\partial}{\partial\theta}-\Omega\right)^{2}+2\pi\Tilde{\gamma}\frac{\abs{\varphi(\theta)}^{2}}{\eta(\theta)^{2}} \\
&+\frac{1}{\Tilde{a}_{\perp}^{2}}\left(\frac{1}{\eta(\theta)^{2}}+\eta(\theta)^{2}\right)\Bigg]\varphi(\theta) ,
\eal
\label{1dNPSEdimless}
\eeq
with
\beq
\Tilde{\gamma}\equiv \frac{2}{\pi}\frac{\Tilde{a}_{s}}{\Tilde{a}_{\perp}^{2}}, \quad\quad
\eta(\theta)=\left(1+2\Tilde{a}_{s}\abs{\varphi(\theta)}^{2}\right)^{1/4}. 
\label{gameta}
\eeq
Here we have introduced the dimensionless $s$-wave scattering length $\Tilde{a}_{s}\equiv Na_{s}/R$ and the dimensionless transverse width $\Tilde{a}_{\perp}\equiv a_{\perp}/R$. 

\begin{figure}[t]
\centering
\includegraphics[width=80mm]{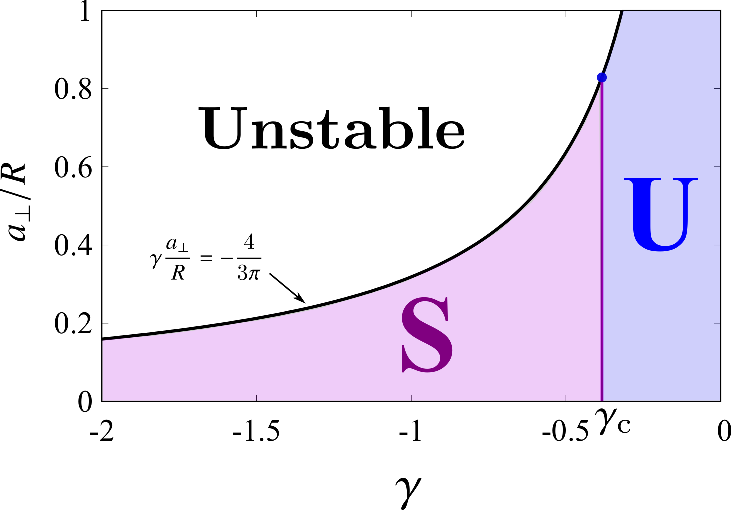}
\caption{Phase diagram for the density configuration. The symbols ``{\bf S}'' and ``{\bf U}'' stand for solitonic regime and uniform regime respectively, which are separated by the critical interaction strength $\gamma_{\mathrm{c}}=2\omega^{2}-1/2$. For $\gamma\Tilde{a}_{\perp}<-4/(3\pi)$, the system undergoes a dynamical instability. }
\label{bsphase}
\end{figure}

In the thin transverse length limit $a_{\perp}\to0$ with a fixed interaction strength $\Tilde{\gamma}$, one finds $\eta(\theta)\to1$ and the 1D NPS equation in Eq.~\eqref{1dNPSEdimless} recovers the 1D GP equation as
\beq
\bar{\mu}_{\mathrm{eff}}\varphi(\theta)
=\left[\left(-i\frac{\partial}{\partial\theta}-\Omega\right)^{2}+2\pi\Tilde{\gamma}\abs{\varphi(\theta)}^{2}\right]\varphi(\theta) ,
\label{npseGP}
\eeq
with $\bar{\mu}_{\mathrm{eff}}\equiv\bar{\mu}-2\Tilde{a}_{\perp}^{-2}$ the dimensionless effective chemical potential. 

We illustrated the density configuration in the left panel of Fig.~\ref{densfracap} for $\Tilde{\gamma}=-1$. 
As one increases the transverse width $\Tilde{a}_{\perp}$, we have a sharper bright soliton. 
The right panel in Fig.~\ref{densfracap} illustrates the superfluid fraction without a boost calculated by Leggett's formula in Eq.~\eqref{leggettformula} solving Eq.~\eqref{1dNPSEdimless} numerically. 
We can check that the superfluid fraction approaches the results by the 1D GP equation in $a_{\perp}\to0$, which is consistent with Eq.~\eqref{npseGP}. 
As one increases the transverse width $a_{\perp}/R$, for $\Tilde{\gamma}=-0.5$ and $\Tilde{\gamma}=-0.7$, the superfluid fraction decays up to a certain value of $a_{\perp, \mathrm{c}}/R\equiv\Tilde{a}_{\mathrm{c}}$. 
Beyond $\Tilde{a}_{\mathrm{c}}$, $\eta^{4}(\theta)$ in Eq.~\eqref{gameta} is negative and therefore $\eta^{2}(\theta)$ changes to purely imaginary. 
In this case, one can see $\mathrm{Im}[\mu]>0$, which indicates $\Psi(x,t)\propto e^{\mathrm{Im}[\mu]t/\hbar}$ and the steady-state solution in Eq.~\eqref{steady} is dynamically unstable. 
This is the collapse of the condensate implicitly induced by the transverse dynamics encoded into the NPS equation \eqref{TDNPS}. 
The stability condition of the bright soliton is given by \cite{salasnichb} 
\beq
-\frac{4}{3\pi}<\Tilde{\gamma}\Tilde{a}_{\perp}<0 ,
\eeq
which gives the critical transverse width $\Tilde{a}_{\mathrm{c}}\equiv4/\left(3\pi\abs{\Tilde{\gamma}}\right)$. 
On the other hand, in the case of $\Tilde{\gamma}=-0.3$, the superfluid fraction does not decay at all by increasing the transverse width. 
This is because the density configuration is uniform instead of a bright soliton as in the case of a 1D GP equation below the critical interaction strength $\abs{\gamma}<1/2$. 
The phase diagram for the density configuration is summarized in Fig.~\ref{bsphase}.

\section{Effects of potential barrier}

In this section, we examine the effects of potential barriers in the ring on the density configuration and superfluid fraction. 

\begin{figure}[t]
\centering
\includegraphics[width=80mm]{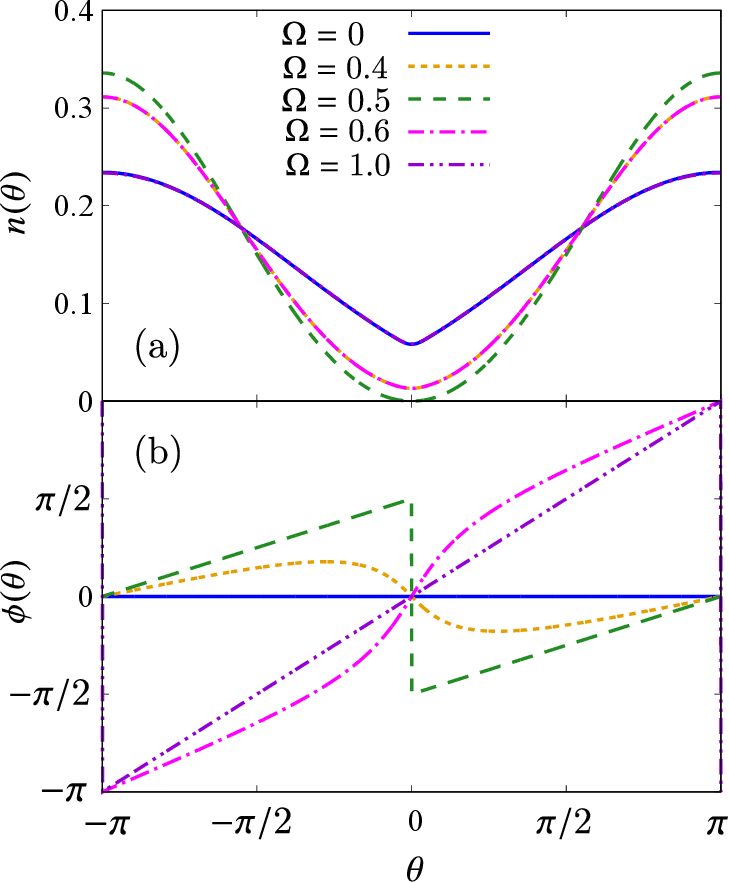}
\caption{Density distribution and phase variation for $\gamma=0.1$ and $w=0.1$. 
The upper panel (a) shows the density distribution $n(\theta)$ while the lower panel (b) shows the phase variation $\phi(\theta)$ with varying boost velocity. }
\label{Figdensityphasebar}
\end{figure}
\begin{figure*}[t]
\begin{tabular}{cc}
\begin{minipage}{0.5\hsize}
\centering
\includegraphics[scale=0.7]{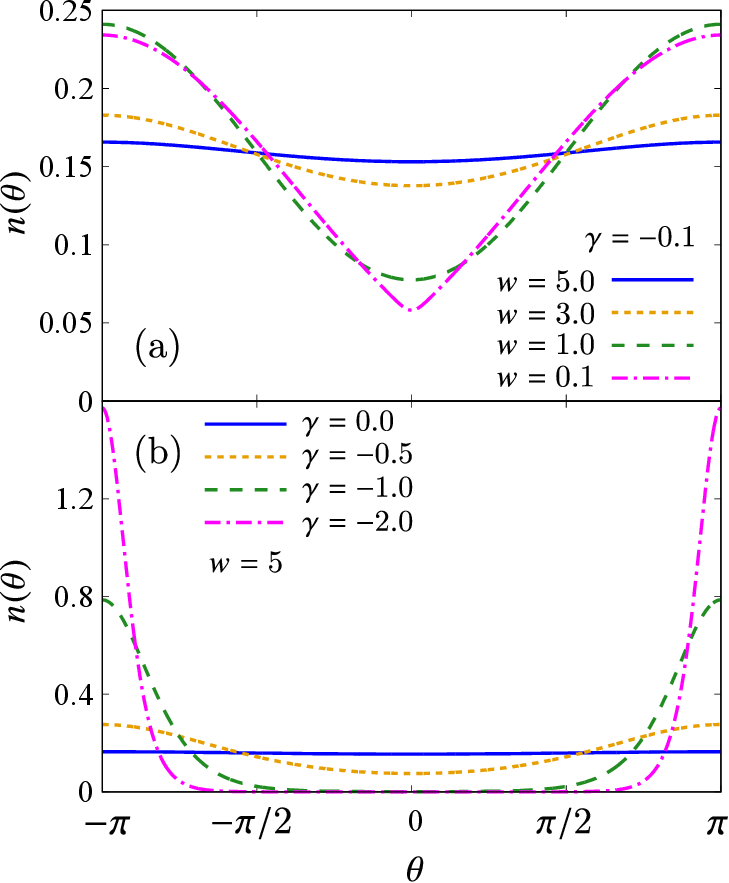}
\end{minipage}
\begin{minipage}{0.5\hsize}
\centering
\includegraphics[scale=0.7]{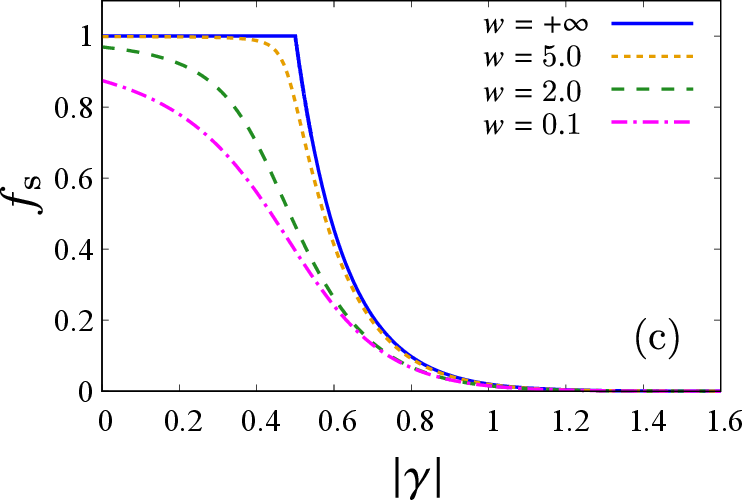}
\end{minipage}
\end{tabular}
\caption{Density distribution $n(\theta)=\abs{\varphi(\theta)}^{2}$ and the superfluid fraction in the presence of a Gaussian barrier numerically calculated by Leggett's formula \eqref{leggettformula} without boost. The upper panel (a) in the left figure shows the density distribution at a fixed interaction strength $\gamma=-0.1$. By decreasing the barrier width, it undergoes modulational transformation and forms a dark soliton around $\theta=0$. The lower panel (b) in the left figure shows the density distribution at a fixed barrier width $w=5$. 
As one increases the attractive interaction strength $\gamma$, the uniform density distribution turns to a dark soliton around $\theta=0$. 
In the right panel (c), the blue solid line corresponds to the superfluid fraction without the barrier $w=+\infty$. A finite barrier width makes the modulation nonuniform and the superfluid fraction deviates from $f_{\mathrm{s}}=1$. In addition, the abrupt decrease at the critical interaction strength is smeared.}
\label{densfracGauss}
\end{figure*}
\subsection{Single Gaussian barrier}\label{SecGauss}

\begin{figure*}[t]
\centering
\includegraphics[width=80mm]{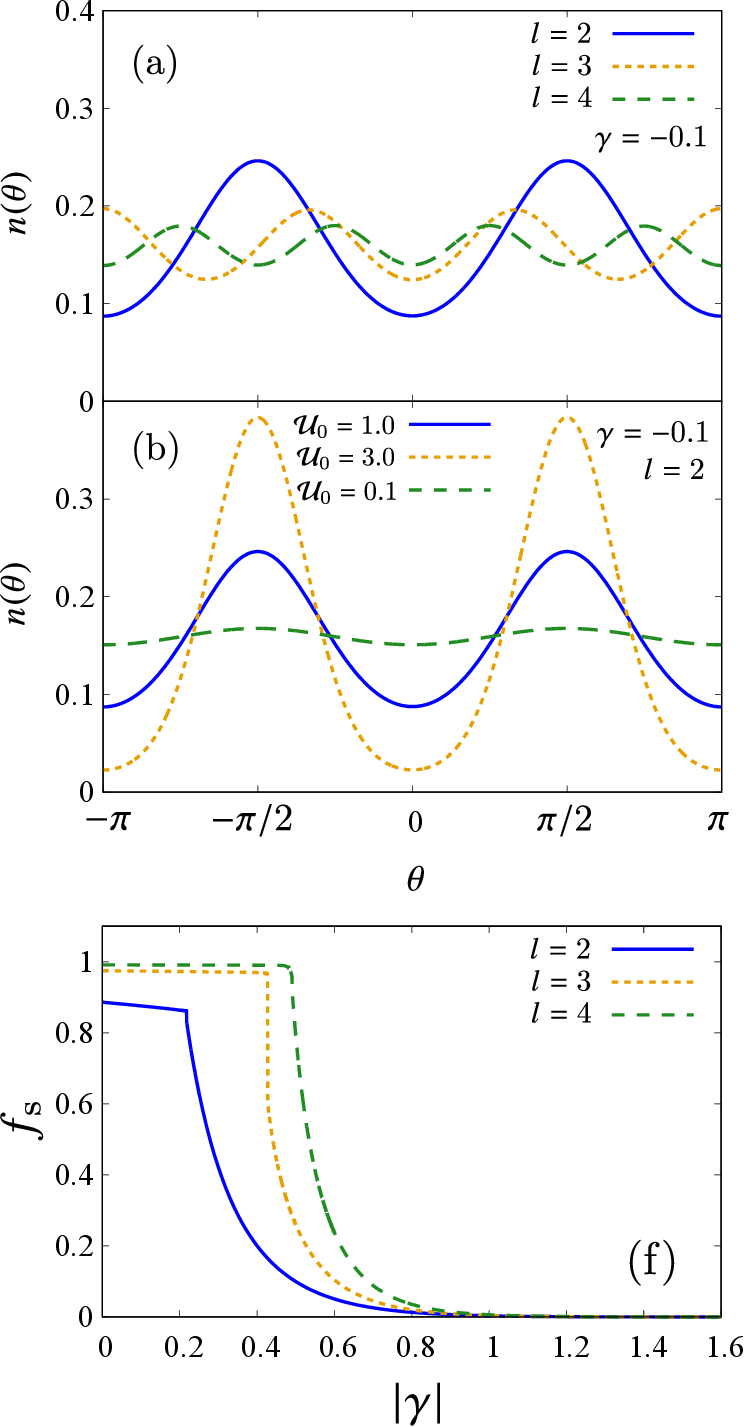}
\includegraphics[width=80mm]{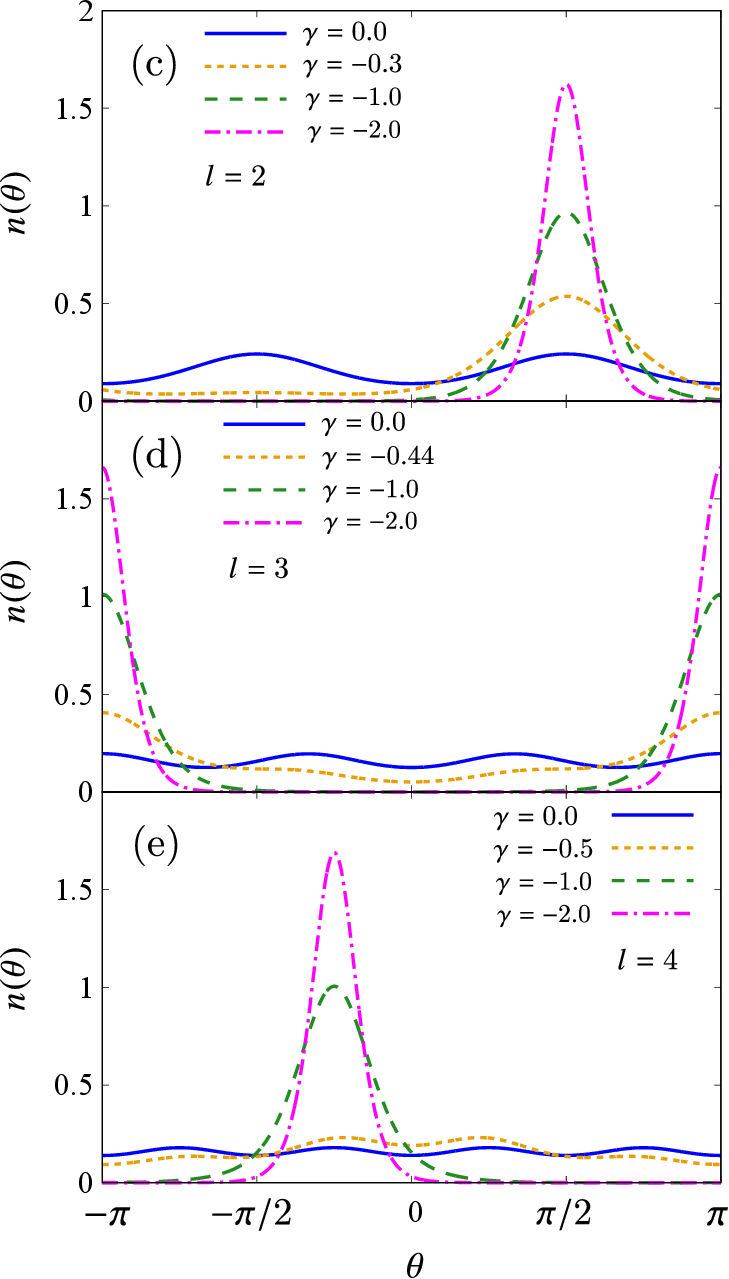}
\caption{Density distribution $n(\theta)=\abs{\varphi(\theta)}^{2}$ and the superfluid fraction $f_{s}$ without boost under a spatially periodic potential $\mathcal{U}_{l}(\theta)=\mathcal{U}_{0}\cos{\left(l\theta\right)}$. 
The upper left panel (a) displays the density configuration for $\gamma=-0.1$ under $\mathcal{U}_{0}=1$. 
The blue solid curve, the orange dotted curve, and the green dashed curve stand for the results of $l=2,3,4$ respectively. 
Panel (b) shows the density with some different values of potential amplitude $\mathcal{U}_{0}=1.0, 3.0, 0.1$ for $\gamma=-0.1$ and $l=2$. 
Each of the right panels (c), (d), (e) represent the density distributions for $l=2,3,4$, respectively, with varying interaction strength under $\mathcal{U}_{0}=1$. 
The lower left panel (f) shows the superfluid fraction calculated by Leggett's formula \eqref{leggettformula} under $\mathcal{U}_{0}=1$. 
The blue solid, orange dotted, and green dashed lines stand for the results for $l=2,3,4$, respectively. }
\label{denbarcos}
\end{figure*}

As suggested by Ref.~\cite{minguzzi}, a single potential barrier in the annular superfluid allows us to observe a soliton located at the opposite side of the potential center. 
Let us assume a Gaussian barrier located at origin
\beq
U(x)=\frac{\hbar^{2}}{2mR^{2}}\mathcal{U}(\theta)
=\frac{\hbar^{2}}{2mR^{2}}\frac{\mathcal{U}_{0}}{\sqrt{\pi}w}e^{-\theta^{2}/w^{2}},
\eeq
with $w$ the dimensionless barrier width. 
In the thin barrier limit $w\to0$, it approaches $\mathcal{U}(\theta)\to\mathcal{U}_{0}\delta(\theta)$. 
In $w\to\infty$, on the other hand, $\mathcal{U}(\theta)\to0$ and it recovers a barrier-free annular superfluid analyzed in the previous section. 
We set $\mathcal{U}_{0}=1$ for simplicity. 

We illustrate the density distribution and the phase variation numerically calculated from the GP equation for $\gamma=-0.1$ and $w=0.1$ in Fig.~\ref{Figdensityphasebar}. 
The density in panel (a) exhibits a periodicity in $\Omega$ and involves a node only for $\Omega=0.5$ as explained in Sec.~\ref{Sec2}. 
In panel (b), one can observe that it exhibits a node of phase at $\theta=0$ with a finite angular frequency of rotation $\Omega\neq0$. 
At $\Omega=0.5$, in particular, it involves a $\pi$-phase slip at $\theta=0$. 
This means that the solution is a black soliton by definition. 

The left panels in Fig.~\ref{densfracGauss} illustrate the density distribution with some different values of barrier width. 
The upper panel (a) shows the result with a fixed interaction strength $\gamma=-0.1$ under $\Omega=0$. 
With a large barrier width, it is almost uniform distribution because $\abs{\gamma}<\abs{\gamma_{\mathrm{c}}}$. 
As one decreases the barrier width, a suppressed soliton around $\theta=0$ forms. 
In particular, because $\mathcal{U}(\theta)\simeq\mathcal{U}_{0}\delta(\theta)$ with a small $w$, the density $n(\theta)$ exhibits an abrupt change at $\theta=0$.  

To see the modulational changes, the superfluid fraction given by Leggett's formula is useful. 
We show the superfluid fraction in the right panel of Fig. \ref{densfracGauss}. 
The finite barrier width $w$ makes the abrupt decrease at the critical interaction strength smeared and the superfluid fraction deviates from $f_{\mathrm{s}}=1$. 
However, as mentioned, the nonuniform modulation that leads to $f_{\mathrm{s}}<1$ is crucially different between the case of a finite barrier width and that of $w=+\infty$. 
In the former case, we have a bright soliton at $\theta=0$ for $\gamma<\gamma_{\mathrm{c}}$. 
In the latter case, on the other hand, the bright soliton is shifted to $\theta=\pi$ because of the potential barrier at $\theta=0$ at any $\gamma<\gamma_{\mathrm{c}}$. 

\subsection{Spatially periodic potential}

Here we consider the effects of a spatially periodic potential
\beq
U_{l}(x)=\frac{\hbar^{2}}{2mR^{2}}\mathcal{U}_{l}(\theta)
=\frac{\hbar^{2}}{2mR^{2}}\mathcal{U}_{0}\cos{\left(l\theta\right)}, 
\eeq
with $l$ an integer. 
In particular, $\mathcal{U}_{2}(\theta)$ corresponds to a double-well potential which has two valleys along the ring \cite{fattori,malomed,reatto}. 
Such a double-well potential is realized also by deforming the ring into an ellipse as a consequence of the geometric potential $U_{\text{geo}}(s)=-\hbar^{2}\kappa(s)^{2}/(8m)$, where $\kappa(s)$ is the curvature with the curvilinear abscissa $s$ along the geometry \cite{leboeuf,schwartz,sandin,moller}. 
In this way, the geometric potential $U_{\text{geo}}$ takes its minima at a point with a maximum curvature, while it takes the maxima at a point with a minimum curvature. 
Along an ellipse, consequently, the geometric potential plays the role of a double-well potential. 
This periodic external potential explicitly breaks the continuous translational symmetry. 
Let us see how a soliton forms under such a potential barrier. 

Figure \ref{denbarcos}(a) shows the density configuration for $\gamma=-0.1$ with $l=2,3,4$. 
We set $\mathcal{U}_{0}=1$ as in Sec.~\ref{SecGauss}. 
The uniform configuration in the absence of a potential barrier alters to a lattice structure, which has $l$ maxima, under the spatially periodic barrier $\mathcal{U}_{l}$. 
The amplitude of the density modulation is smaller with a larger integer $l$ indicating a shallow regime. 
In Fig.~\ref{denbarcos}(b), we showed the density distribution with some different values of potential amplitude $\mathcal{U}_{0}=0.1, 1, 3$ for $\gamma=-0.1$ and $l=2$. 
Increasing the potential height $\mathcal{U}_{0}$, we have a highly localized density configuration, while it delocalizes with a small $\mathcal{U}_{0}$. 
Figures \ref{denbarcos}(c)-\ref{denbarcos}(e) illustrate the density configurations with varying interaction strength and $\mathcal{U}_{0}=1$. 
As one increases the attractive interaction strength $\gamma$, the system selects one of the potential minima to develop a bright soliton, and the noninteracting lattice structure transforms to a highly localized configuration. 

We show the superfluid fraction in Fig.~\ref{denbarcos}(f) for $l=2,3,4$ with $\mathcal{U}_{0}=1$. 
The formation of lattice configuration in the weakly interacting regime leads to the deviation of the superfluid fraction from $f_{\mathrm{s}}=1$. 
We can see that the critical interaction strength at which the superfluid fraction abruptly drops decreases with a small $l$. 
It indicates that the formation of a bright soliton starts to occur with a smaller interaction strength. 
As one increases $l$, the behavior of the superfluid fraction approaches that without an external potential because increasing $l$ corresponds to an almost uniform potential which can be negligible.

\section{Conclusion}
We elucidated superfluid properties of a one-dimensional annular superfluid, which exhibits solitons with sufficiently attractive interaction, subject to a rotational boost, a transverse dimension, or a potential barrier by using the superfluid fraction given by Leggett's formula. 
We derived Leggett's formula for the superfluid fraction, which does not contradict Landau's criterion and is valid even in the noninteracting case with a finite excitation gap in a finite-size system. 
The superfluid fraction is $f_{\mathrm{s}}=1$ for a uniform configuration and decreases for an inhomogeneous density. 
Therefore, Leggett's formula characterizes modulational instabilities and we have seen the formation of soliton abruptly reduces the superfluid fraction. 
In particular, at $\omega=-1/2$, we have a black soliton and the superfluid fraction vanishes. 
We have included effects of transverse width through the one-dimensional nonpolynomial Schr$\ddot{\text{o}}$dinger equation. 
The superfluid fraction monotonically decreases as one increases the transverse width up to the critical width, above which the superfluid is unstable. 
We have also investigated the superfluid fraction in the presence of a potential barrier. 
The potential barrier alters a uniform configuration without a barrier into an inhomogeneous one, which reduces the superfluid fraction. 
As one increases the attractive interaction strength, the system selects one of the potential minima to develop a soliton. 
In a dipolar supersolid, the superfluid fraction associated with the nonclassical rotational inertia has been experimentally observed through the measurement of the scissors mode frequency \cite{tanzi}. 
We expect that our results of the superfluid fraction in an annular superfluid are also experimentally accessible in a similar manner. 
A one-dimensional superfluid is a simple and useful platform to analyze various macroscopic quantum phenomena such as Josephson dynamics \cite{bouchoule,tononi20,binanti,vinh}. 
Moreover, the formation of solitons that we have focused on appears ubiquitously not only in condensed-matter physics \cite{togawa,kishine} but in high-energy physics \cite{nitta,brauner}. 
Our work on the superfluid properties of the one-dimensional annular superfluid would have a wide range of applications and contributions to further theoretical investigations in diverse physical systems. 

\begin{acknowledgments}
We thank F. Ancilotto for fruitful discussions and L. Amico for useful comments. 
K.F. is supported by a Ph.D. fellowship of the Fondazione Cassa di Risparmio di Padova e Rovigo. 
\end{acknowledgments}

\end{document}